\documentclass[twocolumn,prb,aps,10pt]{revtex4-1}
\usepackage{graphicx,epsf}

\begin{document}

\title{Probing nonlinear mechanical effects through electronic
currents: \\the case of a nanomechanical resonator acting as
electronic transistor}

\author{A. Nocera$^{1,2}$, C.A. Perroni$^{3,4}$, V. Marigliano Ramaglia$^{1,4}$ and V. Cataudella$^{3,4}$ }

\affiliation{$^{1}$CNISM, $^{2}$Dipartimento di Fisica E. Amaldi, Universita' di Roma Tre, Via della Vasca Navale 84, I-00146 Roma, Italy\\
$^{3}$CNR-SPIN, $^{4}$Universita' degli Studi di Napoli Federico
II, Complesso Universitario Monte Sant'Angelo, Via Cintia, I-80126
Napoli, Italy.}

\begin{abstract}
We study a general model describing a self-detecting single
electron transistor realized by a suspended carbon nanotube
actuated by a nearby antenna. The main features of the device,
recently observed in a number of experiments, are accurately
reproduced. When the device is in a low current-carrying state, a
peak in the current signals a mechanical resonance. On the
contrary, a dip in the current is found in high current-carrying
states. In the nonlinear vibration regime of the resonator, we are
able to reproduce quantitatively the characteristic asymmetric
shape of the current-frequency curves. We show that the nonlinear
effects coming out at high values of the antenna amplitude are
related to the effective nonlinear force induced by the electronic
flow. The interplay between electronic and mechanical degrees of
freedom is understood in terms of an unifying model including in
an intrinsic way the nonlinear effects driven by the external
probe.
\end{abstract}

\maketitle
\newpage

\section{Introduction}

It has been recently shown that carbon nanotubes can act
simultaneusly as single electron transistors \cite{Witkamp} (SET)
and as nanoeletromechanical systems (NEMS) \cite{Craighead,
Ekinci}. The idea is to use a single carbon nanotube placed
between two metal contacts in a suspended configuration as a
self-detecting SET. Due to the extreme properties of carbon
nanotubes (ideal for NEMS applications, because they have a low
mass and a high Young's modulus), the electronic current flowing
through the device results very sensitive to the dynamics of the
nanotube itself.

Although the main effort has been focused on detecting the quantum
regime of mechanical resonators \cite{LaHaye,RoukesB,Connell},
recently, G. A. Steele \emph{et al.}\cite{Steele} and A. K. Huttel
\emph{et al.}\cite{Huttel} were able to fabricate a carbon
nanotube electromechanical device working in the semiclassical
regime (resonator frequencies in MHz range compared to an
electronic hopping frequency of the order of tens of GHz) with an
extremely large quality factor ($Q>10^{5}$). By measuring the
variations of the electronic current flowing through the nanotube
as function of the frequency of a nearby antenna actuating its
motion, they were able to detect very well defined resonances
corresponding to the bending mode of the nanotube itself. The
possibility of using currents to probe nanomechanical resonances
is strongly related to the extremely large quality factor
obtained. In particular, when the SET is in a low current-carrying
state (far from electronic resonance), a peak in the current
signals a mechanical resonance, while a dip is observed in a high
current-carrying state (SET in electronic resonance). Moreover, by
adjusting the antenna power, the nanotube resonator can easily be
tuned into the nonlinear vibration regime. For small antenna
amplitudes, the current-frequency curves are very well fitted by a
Lorentzian, while a characteristic triangular shape or even
hysteresis for large antenna amplitudes is obtained.
Interestingly, the operating temperature can affect the
nonlinearity and the quality factor of the resonator in a non
expected way: the nonlinear effects in current-frequency curves
are washed out increasing the temperature \cite{Nota0}. A detailed
analysis of a single frequency resonance dip has also shown that a
broadening can be obtained increasing the source-drain voltage
\cite{Steele}. In Ref. \onlinecite{Steele}, the authors were able
to provide an explanation of some of the observed effects in terms
of a model in which the gate voltage acquires an assigned time
dependence. Within this phenomenological model, the back-action of
the nanotube motion on detected current is understood neglecting
completely the dynamics of the resonator and analyzing the problem
directly at mechanical resonance conditions only in the limit of
small external antenna amplitudes (linear response regime).

In this paper, we do not consider a priori assumptions on the
resonator (nanotube) dynamics. Actually, in the device
investigated in Ref. \onlinecite{Steele,Huttel}, the chemical
potentials of the leads, to which the nanotube is anchored, differ
by the value of the applied transport voltage $eV_{bias}$ (where
$e$ is the electron charge), and thus the environment that the
nanotube experiences cannot be considered at equilibrium since the
voltages applied in the experiment are typically greater than the
temperature, $eV_{bias}>>k_{B}T$ ($k_{B}$ being Boltzmann
constant). Therefore, in order to understand the behavior of the
device under such conditions, one needs to determine
self-consistently the influence of electrical current and of the
external antenna on the resonator dynamics, and vice versa, the
influence of nonthermal nanotube vibrations driven by the antenna
on the electronic current \cite{Nota000}. As rigorously
demonstrated by Mozyrsky \emph{et al.} \cite{Mart}, and reobtained
by us in a different way \cite{Nocera}, in absence of the external
antenna, the vibrational dynamics of the nanotube can be
described, employing a separation between slow vibrational and
fast electronic time scales \cite{Nocera,Faz,Bode1,Bode2}, by a
Langevin equation
\cite{Pisto,Brandes,Brandes1,Bode1,Bode2,Weick1}. This equation is
ruled by an effective force as well as a damping and diffusive
terms stemming from the interaction of the resonator with the
electronic bath consisting of both the nanotube itself (that can
be described by a single electronic level \cite{Weick1}) and the
out-of-equilibrium environment given by the macroscopic leads. By
including the external antenna effects through a forcing term in
the Langevin equation, we provide such a self-consistent
description of vibrational and electronic dynamics as argued
above.

A theoretical treatment of the nanotube based device investigated
in \onlinecite{Steele,Huttel}, including the external antenna
effects, has been already considered by  G. Labadze and Ya. M.
Blanter \cite{Labadze}. In Ref. \onlinecite{Labadze}, the authors
use a Fokker-Plank equation for the resonator distribution
probability based on master
equations\cite{Armour,Blanter,Blanter1,Bennett,Weick}. This
approach implicitely assumes that the energy scale of the applied
voltages is much larger than the electronic tunneling energy scale
$\hbar\Gamma$, $eV_{bias}>>\hbar\Gamma$. Moreover, tunneling is
considered in the \emph{sequential} regime and quantum effects in
the electronic dynamics such as \emph{cotunneling} are disregarded
\cite{Nota00}. We point out that these effects can be important to
interpret the experimental results obtained by G. A. Steele
\emph{et al.} \cite{Steele} and A. K. Huttel \emph{et al.}
\cite{Huttel} since, as already emphasized by G. Weick \emph{et
al.}\cite{Weick1}, the temperature is much smaller than
$\hbar\Gamma$ and the bias voltage \emph{effectively} applied to
the electronic level of the nanotube can be less than or of the
same order of magnitude of the electronic tunneling energy,
$k_{B}T << eV^{eff}_{bias} \leq \hbar\Gamma$ \cite{Nota2}. Our
approach, based on an adiabatic expansion of the time-dependent
electronic Green function on the Keldysh contour in the small
parameter $\omega_{0}/\Gamma$ \cite{Nocera,Faz,Bode1,Bode2}, takes
into account from the beginning of all higher order terms in the
tunneling matrix element between the leads and the nanotube (so
sequential tunneling and cotunneling regimes are described in a
unified way).

The inclusion of the external antenna effects in the effective
Langevin equation for the resonator has allowed us to explore also
the nonlinear renspose regime in a nonperturbative way. We point
out that, even describing the vibrational dynamics of the nanotube
as a single \emph{harmonic} vibrational mode, we are able to
reproduce all the main features observed in Ref.
\onlinecite{Steele,Huttel} including the nonlinear effects.
Indeed, in the nonlinear vibration regime of the resonator, we are
able to reproduce quantitatively the characteristic asymmetric
shape of the current-frequency curves. The observed nonlinearity
is due to the intrinsic nonlinear terms of the effective force
stemming from the strong interaction between electron and
resonator dynamics \cite{Mart,Pisto,Brandes,Brandes1,Nocera}.
These terms are completely neglected in Ref. \onlinecite{Labadze},
loosing any possibility to describe renormalization frequency
effects and to explore the nonlinear response regime for the
resonator. We show that nonlinear effects can be highlighted
\emph{only dynamically} by applying large external antenna
amplitudes. In this sense, our approach is different from other
theoretical studies where nonlinear terms are discussed only
statically \cite{Steele} or are added from the beginning assuming
that the resonator is characterized by anharmonic
terms.\cite{Weick,Weick1,Rama}

Within our approach, the experimental results obtained in Ref.
\onlinecite{Steele,Huttel} in the linear response regime are also
reproduced, as well as the broadening of the mechanical resonance
dip as function of the applied bias voltage. Furthermore, we are
able to predict, in the limit of large bias, the onset of a fine
double dip structure that could be experimentally observed.

The paper is organized as follows: In Sec. II we present the model
able to describe the electronic transistor consisting of the
vibrating nanotube. The equation of motion describing the
resonator dynamics including the external antenna effects is also
discussed. In Sec. III we present numerical results.

\section{Model and method}

We describe the suspended carbon nanotube with a single impurity
Holstein model \cite{Hol}, that is able to catch the main physical
ingredients of the experiments in Ref. \onlinecite{Steele,Huttel}.
For the sake of clarity, we here point out that the model used in
\cite{Steele,Huttel} and in other papers in the literature
\cite{Labadze,Blanter,Blanter1,Bennett,Weick,genovesi}, based on
on a capacitive coupling of the nanotube (dot) to the gate
electrode, is equivalent to a Holstein-like coupling between the
occupation on the dot and the vibrational degree of freedom (see
Appendix A).

As suggested by Weick \emph{et al.} [\onlinecite{Weick}], in the
small energy window of interest for a single dip feature, the
electronic part of the device is modeled as a single electronic
level coupled to the leads through standard tunneling terms.

The electronic Hamiltonian is given by
\begin{equation}\label{Hel}
\hat{\cal H}_{el}=V_{gate}^{eff}{\hat d^{\dag}}{\hat
d}+\sum_{k,\alpha}(V_{k,\alpha}{\hat c^{\dag}_{k,\alpha}}{\hat d}+
h.c.)+ \sum_{k,\alpha}\varepsilon_{k,\alpha}{\hat
c^{\dag}_{k,\alpha}}{\hat c_{k,\alpha}},
\end{equation}
where nanotube's electronic level has energy $V_{gate}^{eff}$ with
creation (annihilation) operators ${\hat d^{\dag}} ({\hat d})$
\cite{Nota1}. The operators ${\hat c^{\dag}_{k,\alpha}} ({\hat
c}_{k,\alpha})$ create (annihilate) electrons with momentum $k$
and energy $\varepsilon_{k,\alpha}=E_{k,\alpha}-\mu_{\alpha}$ in
the left ($\alpha=L$) or right ($\alpha=R$) free metallic leads,
while the electronic tunneling between the molecular level and a
state in the lead has amplitude $V_{k,\alpha}$. The chemical
potentials in the leads $\mu_{L}$ and $\mu_{R}$ are assumed to be
biased by an external voltage $eV_{bias}^{eff}=\mu_{L}-\mu_{R}$.
The coupling to the leads is described by the tunneling rate
$\Gamma_{\alpha,k}=2\pi\rho_{\alpha}|V_{k,\alpha}|^{2}/\hbar$,
where $\rho_{\alpha}$ is the density of states in the lead
$\alpha$. We will suppose symmetric coupling
($\Gamma_{L,k}=\Gamma_{R,k}$) and a flat density of states for the
leads, considered as thermostats at finite temperature, within the
wide-band approximation ($\Gamma_{\alpha,k}\mapsto
\Gamma_{\alpha}$, $\alpha=L,R$) \cite{Weick,Weick1}.

The Hamiltonian of the mechanical degree of freedom is given by
$\hat H_{osc}={\hat p^{2}\over 2m} + {1\over 2}m
\omega^{2}_{0}\hat x^{2}$, characterized by the frequency
$\omega_{0}$ and the effective mass $m$ ($k=m\omega_{0}^{2}$). The
interaction is provided by ${\hat H}_{int}=\lambda \hat x{\hat
N_{el}}$ \cite{Weick,Armour}, where $\lambda$ is the
electron-oscillator coupling strength and ${\hat N_{el}}=\hat
d^{\dag}d$ represents the electronic occupation on the nanotube
(see also Appendix A). Definitely, the overall Hamiltonian is
\begin{equation}\label{Htot}
\hat{\cal H}=\hat{\cal H}_{el}+ \hat H_{osc} +{\hat H}_{int}.
\end{equation}
In this paper, we will measure lengths in units of
$x_{0}={\lambda\over k}$ and energies in units of $\hbar\Gamma$.

The experimental values of the resonance frequencies of the
vibrating nanotube (120-300 MHz range) suggest that the
vibrational motion is very slow compared to the electronic
tunneling rate on the nanotube itself (adiabatic limit):
$\omega_{0}/\Gamma <<1$. Moreover, it was estimated in
Ref.[\onlinecite{Weick}] that for the experiment in consideration
the coupling energy describing electron-phonon interaction,
$E_{p}={\lambda^2\over 2k}\simeq5\mu e V$, implying a strong
coupling between the electronic and vibrational degrees of freedom
($E_{p}/\hbar\omega_{0}=10$). Summarizing, the regime of the
relevant parameters is $\hbar\omega_{0}<<E_{p} (\sim k_{B}T) <<
eV^{eff}_{bias} \leq \hbar\Gamma$ \cite{Nota2}.

As discussed in Ref.[\onlinecite{Nocera}] and
[\onlinecite{Millis}], when $eV_{bias}^{eff}>> \hbar\omega_{0}$
and $k_{B}T > \hbar\omega_{0}$, the semi-classical treatment of
the oscillator dynamics is well justified. Within a
non-equilibrium adiabatic approximation
\cite{Mart,Pisto,Brandes,Nocera,Faz}, the vibrational dynamics of
the nanotube can be described by a \emph{nonstandard Langevin
equation} controlled by a self-consistent effective
\emph{anharmonic} force as well as by damping and diffusive terms
depending explicitely on the resonator displacement $x$. The
oscillator dynamics can be obtained by solving numerically the
following equation
\begin{eqnarray}\label{Langevin1}
m \ddot{x} &+& A(x)\dot{x}=F(x)+\xi(t)+
A_{ext}\cos(\omega_{ext}t),
\end{eqnarray}
where $A_{ext}$ and $\omega_{ext}$ represent the amplitude and the
external antenna frequency, respectively. Furthermore, in
eq.(\ref{Langevin1}), the position-dependent force $F(x)$, damping
$A(x)$, and the intensity of the noise $D(x)$,
$\langle\xi(t)\xi(t')\rangle=D(x)\delta(t-t')$ (where $\xi(t)$ is
a standard white noise term), are related to the electronic
Green's functions on the Keldysh contour as
\cite{Mart,Pisto,Brandes,Nocera},
\begin{eqnarray}
F(x)&=&-k x+\lambda\langle \hat N_{el}\rangle(x),\label{for}\\
\langle \hat N_{el}\rangle(x)&=&\int
{d\hbar\omega\over2\pi\imath}G^{<}(\omega,x)\\
A(x)&=&\lambda^{2}\int
{d\hbar\omega\over2\pi}G^{<}(\omega,x)\partial_{\hbar\omega}G^{>}(\omega,x),\\
D(x)&=&\lambda^{2}\int
{d\hbar\omega\over2\pi}G^{<}(\omega,x)G^{>}(\omega,x).
\end{eqnarray}
The lesser $G^{<}$ and greater $G^{>}$ Green's functions at finite
temperature are given by
\begin{eqnarray}\label{GlesGgrt}
G^{<}(\omega,x)&=&{\imath\hbar\Gamma\over 2}{
f_{L}(\omega)+f_{R}(\omega)\over
(\hbar\omega-V_{gate}^{eff}-\lambda x)^{2}+
(\hbar\Gamma/2)^{2}},\\
G^{>}(\omega,x)&=&{-\imath\hbar\Gamma\over 2}{
2-f_{L}(\omega)-f_{R}(\omega)\over
(\hbar\omega-V_{gate}^{eff}-\lambda x)^{2}+ (\hbar\Gamma/2)^{2}},
\end{eqnarray}
where $f_{L,R}(\omega)$ are the Fermi functions of the leads and
we have assumed that
$\hbar\Gamma_{L}=\hbar\Gamma_{R}=\hbar\Gamma/2$. At low
temperatures, $k_{B}T<<\hbar\Gamma$, the Fermi function
$f_{\alpha}(\omega)$ can be replaced by the step function
$\Theta(\hbar\omega-\mu_{\alpha})$ ($\alpha=L,R$), obtaining
\begin{eqnarray}
\langle N_{el}\rangle(x)&=&{1\over 2\pi}
\sum_{\alpha=L,R}\Bigg(\arctan\Big({\mu_{\alpha}-V_{gate}^{eff}-\lambda
x\over \hbar\Gamma/2}\Big)+{\pi\over2}\Bigg),\nonumber\\\label{occu}\\
A(x)&=&{4m\omega_{0}\over\pi}{\hbar\omega_{0}\over\hbar\Gamma}{E_{p}\over\hbar\Gamma}
\sum_{\alpha=R,L}{1\over
\Big[\Big({\mu_{\alpha}-V_{gate}^{eff}-\lambda x\over
\hbar\Gamma/2}\Big)^{2}+1\Big]^{2}},\nonumber\\\label{damp}\\
D(x)&=&{m\omega_{0} E_{p}\over
\pi}{\hbar\omega_{0}\over\hbar\Gamma}\Bigg\{\arctan\Big({\mu_{\alpha}-V_{gate}^{eff}-\lambda
x\over \hbar\Gamma/2}\Big)\nonumber\\
&+&{{\mu_{\alpha}-V_{gate}^{eff}-\lambda x\over
\hbar\Gamma/2}\over \Big[\Big({\mu_{\alpha}-V_{gate}^{eff}-\lambda
x\over
\hbar\Gamma/2}\Big)^{2}+1\Big]}\Bigg\}^{\alpha=L}_{\alpha=R}.\label{fluc}
\end{eqnarray}
The linear elastic force exerted on the oscillator is modified by
a relevant \emph{nonlinear} correction term proportional to the
electronic occupation eq.(\ref{occu}). The strength of the damping
$A(x)$ and diffusive $D(x)$ terms result proportional to the
adiabatic ratio $\omega_{0}/\Gamma$ and therefore one can safely
neglect their spatial dependence. Anyway, we point out that the
diffusive term $D(x)$ vanishes at equilibrium (bias voltage
$V^{eff}_{bias} = 0$). Only on application of finite bias it
becomes different from zero. In the regime of experimental
parameters explored in Refs. \onlinecite{Steele,Huttel},
$\hbar\omega_{0}<<E_{p} (\sim k_{B}T) << eV^{eff}_{bias} \leq
\hbar\Gamma$, the effect of the electronic bath on the resonator
dynamics can be described by an effective temperature proportional
to the bias voltage $k_{B}T_{eff}\simeq eV_{bias}^{eff}/8$
\cite{Mart,Nocera}.

In eq.($\ref{Langevin1}$), we consider the parameter $A_{ext}$
expressed in terms of the natural force unit
$\lambda=\omega_{0}\sqrt(2 m E_{p})$. Assuming a nanotube mass of
$m\sim 10^{-23}kg$, an oscillation frequency of $120$ MHz,
$\lambda$ is of the order of $10^{-16}N$, while the effective
spring constant is $k=10^{-6} N/m$. By solving the second order
stochastic differential equation (\ref{Langevin1}) \cite{Nocera},
we are able to obtain the distribution probabilities of the
displacement $P(x)$. This allows us to calculate any system
property as an average over the distribution probability $P(x)$.
In particular, in order to make contact with experimental results,
we have calculated the average electronic current $\langle
I\rangle$ flowing through the nanotube as $\langle
I\rangle=\int_{-\infty}^{+\infty}dx I(x)P(x)$, where $I(x)$ is the
current at a particular resonator displacement.

\section{Results}

\begin{figure}
\centering
{\includegraphics[width=9.0cm,height=8.0cm,angle=0]{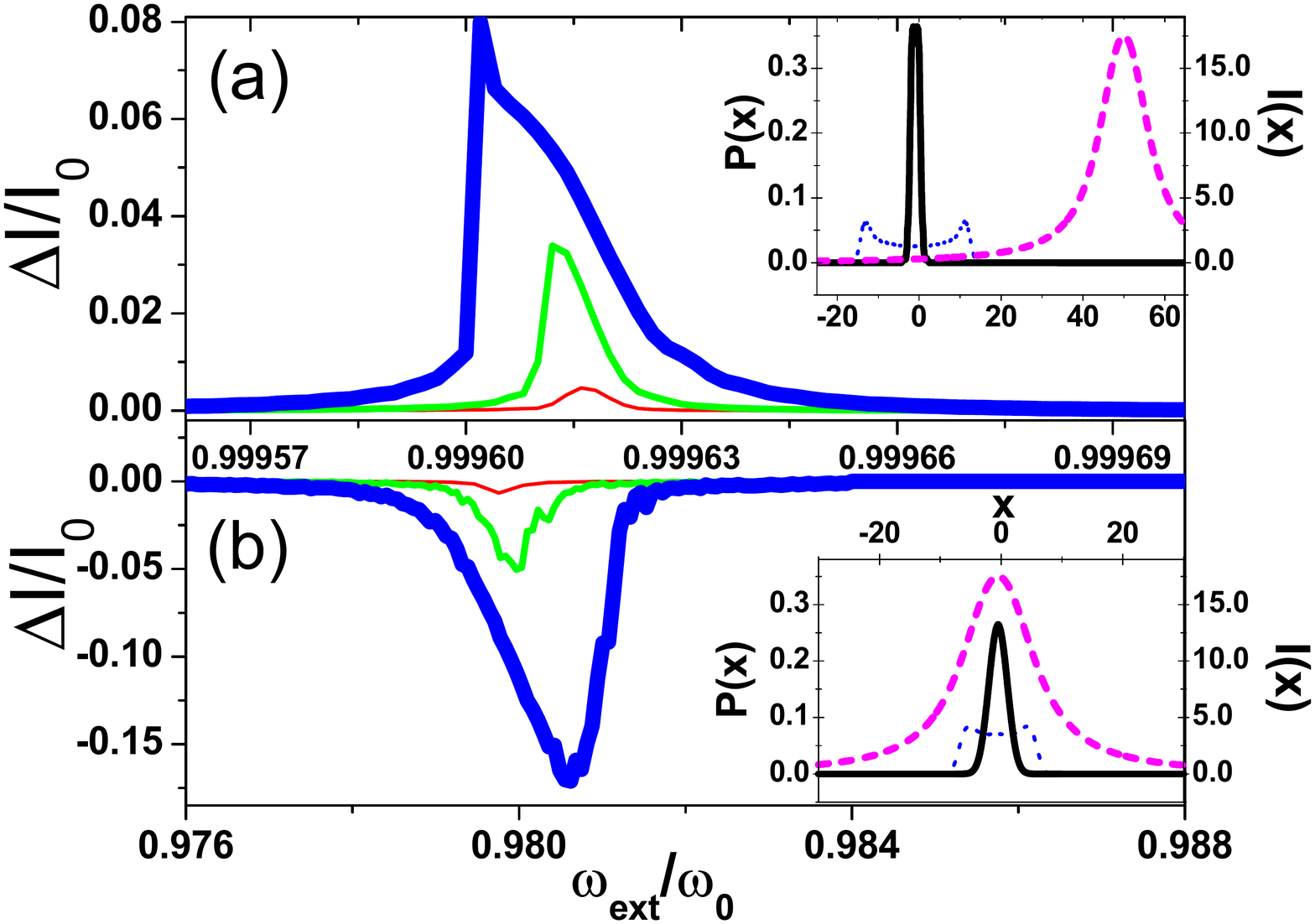}}
\caption{(Color online) Panel(a): Normalized current change
($\Delta I/I_{0}$) in a low current-carrying state
($V_{gate}^{eff}=-4\hbar\Gamma$) as function of the external
frequency ($\omega_{ext}/\omega_{0}$) for different antenna
amplitudes: $A_{ext}=10^{-5}$ solid thin line (red online),
$A_{ext}=10^{-4.5}$ normal thickness line (green online),
$A_{ext}=10^{-3.5}$ thick line (blue online). Panel (b): $\Delta
I/I_{0}$ against $\omega_{ext}/\omega_{0}$ in a high
current-carrying state ($V_{gate}^{eff}=E_{p}$) for different
antenna amplitudes: $A_{ext}=10^{-3.5}$ solid thin line (red
online), $A_{ext}=10^{-3}$ normal thickness line (green online),
$A_{ext}=10^{-2.5}$ thick line (blue online). Insets: solid line
(black online) is a distribution $P(x)$ out of mechanical
resonance. Dotted line (blue online) is a distribution obtained at
mechanical resonance for the larger value of antenna amplitude
considered in main plots of panels (a,b). Short-dashed line
(magenta online) represents current as function of position
$I(x)$. In this plot $eV_{bias}^{eff}=0.5\hbar\Gamma$,
$\omega_{0}/\Gamma=0.004$ and
$E_{p}/\hbar\Gamma=k_{B}T/\hbar\Gamma=0.04$.}\label{fig0}
\end{figure}
One of the main results of Ref. \onlinecite{Steele,Huttel} is the
observation of the electronic current changes at fixed gate
voltage as function of the external antenna frequency and
amplitude. Our model accurately reproduces the experimental
results both in a low current-carrying state, where a peak in the
current signals a mechanical resonance (Fig.\ref{fig0}a), and in a
high current-carrying state, where a dip is observed
(Fig.\ref{fig0}b).

These two behaviors (peak and dip) can be understood considering
that the oscillator explores wider regions in configuration space
when the external antenna amplitude increases (see the
distribution probabilities $P(x)$ shown in the insets of Fig.
\ref{fig0}). When the electronic device is in a low
current-carrying state (Fig.\ref{fig0}a), $P(x)$ is concentrated
at $x$ values far form the configurations where the device carries
the maximum current (inset of Fig.\ref{fig0}a). By increasing the
external antenna amplitude, the resonator is able to explore
larger regions which carry more and more current obtaining a
positive contribution in the normalized electronic current change
$\Delta I/I_{0}$ with respect to the background value $I_{0}$. On
the contrary, when the device is in a high current-carrying state,
the distribution probabilities and the current are centered at the
same position (inset of Fig.\ref{fig0}b). In this case, the effect
of the external antenna is to give a negative contribution in the
normalized electronic current change $\Delta I/I_{0}$ (Fig
\ref{fig0}b) since, the resonator explores regions of phase space
which carry less and less current (inset of Fig. \ref{fig0}b).

The results shown in Fig.\ref{fig0} allow also to characterize the
behavior of the resonator in the nonlinear regime. Interestingly,
(see Fig.\ref{fig0}a), increasing the amplitude of the external
forcing, the shape of the current-frequency curves changes. For
small antenna amplitudes, a characteristic Lorentzian shape is
observed. This is expected for an harmonic oscillator driven by a
periodic forcing in the absence of external noise. In fact, even
if in mechanical resonance, the oscillator explores only small
regions around the stationary point and nonlinear corrections
terms of the force $F(x)$ (eq. (\ref{for})) do not come into play.
At mechanical resonance, only when the amplitude of the external
antenna increases, the oscillator explores a larger region in
phase space, where the nonlinear terms of the force acting on the
oscillator cannot be neglected.

Within our approach, for large antenna amplitudes and in the
presence of noise, the current-frequency profiles assume the
experimentally observed characteristic triangular shape
\cite{Huttel}. Furthermore, a softening is observed when the
device is in a low current-carrying state (Fig. \ref{fig0}a),
while an hardening in a high current-carrying state is obtained
(Fig. \ref{fig0}b). This nonlinear behavior can be understood by
analyzing the properties of the force $F(x)$ (eq. \ref{for})
around the stationary point. Softening and hardening behavior of
the resonance frequency are usually related to the sign of the
cubic nonlinear term \cite{Nayfeh}. When the device is in a low
current-carrying state, indeed, the sign of this term is positive,
giving a net softening effect. In a high current-carrying state
($V^{eff}_{gate}=E_{p}$) and for bias values sufficiently small,
the sign of the cubic nonlinear term is negative providing an
hardening.
\begin{figure}
\centering
{\includegraphics[width=9.0cm,height=8.0cm,angle=0]{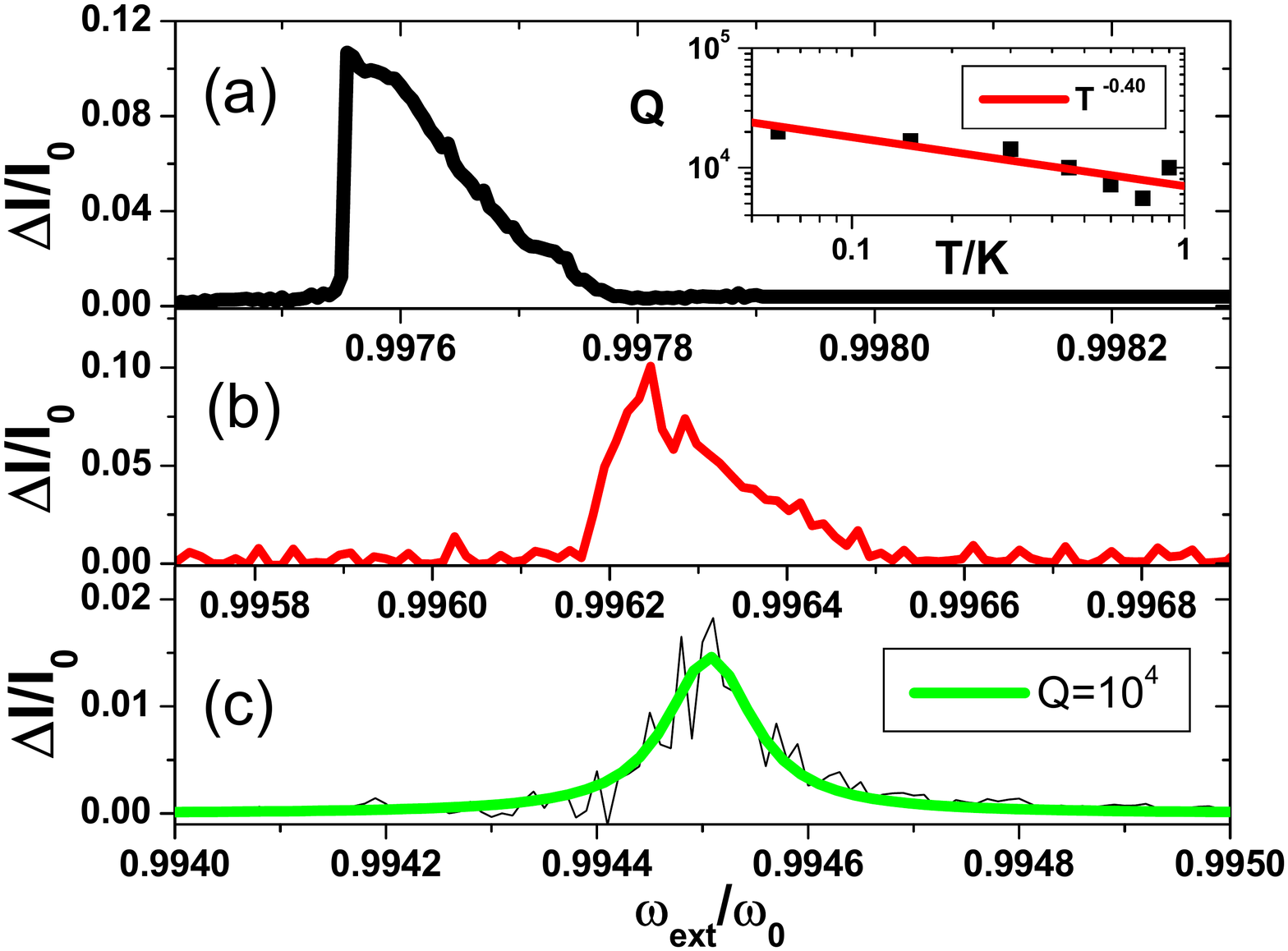}}
\caption{(Color online) $\Delta I/I_{0}$ against
$\omega_{ext}/\omega_{0}$ in a low current-carrying state
($V_{gate}^{eff}=-1.75\hbar\Gamma$) when the resonator is driven
by a strong external antenna amplitude $A_{ext}=10^{-3.5}$: panel
(a) $k_{B}T=0.05\hbar\Gamma$, panel (b) $k_{B}T=0.375\hbar\Gamma$,
panel (c) $k_{B}T=\hbar\Gamma$. In panel (c) a Lorentzian fit is
also drown with $Q=10^{4}$ ($Q$ is defined in the main text).
Inset of panel (a): intrinsic quality factor $Q$ as function of
temperature \cite{Nota3}. We use as energy unit
$\hbar\Gamma=125\mu eV$. In this plot
$eV_{bias}^{eff}=0.5\hbar\Gamma$, $\omega_{0}/\Gamma=0.005$ and
$E_{p}/\hbar\Gamma=0.05$.}\label{fig3}
\end{figure}

\begin{figure}
\centering
{\includegraphics[width=9cm,height=8.0cm,angle=0]{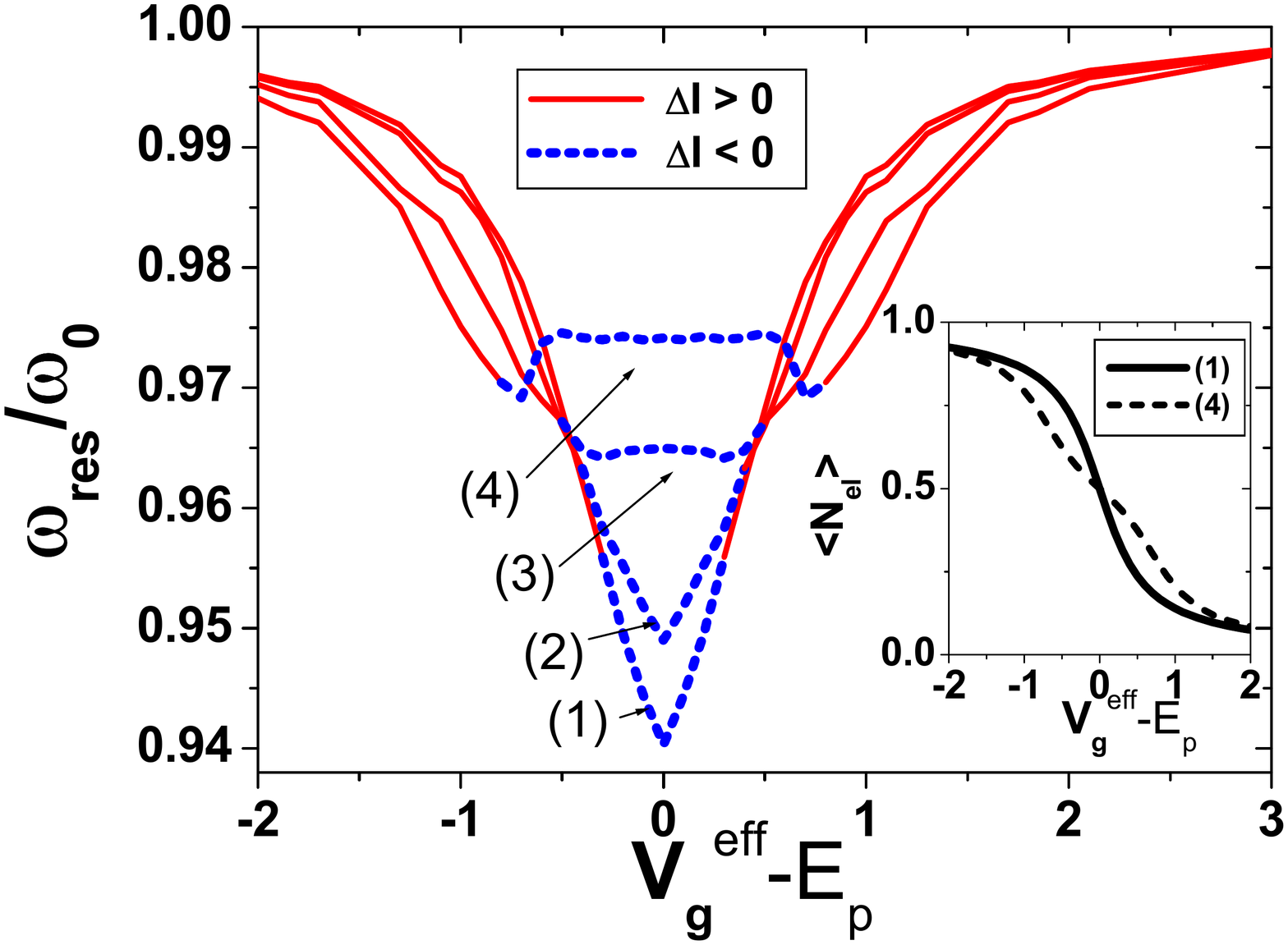}}
\caption{(Color online) Resonator frequency at resonance against
effective gate voltage (shifted of $E_{p}$) for different bias
voltages: $eV_{bias}=0.1\hbar\Gamma$ curve $(1)$,
$eV_{bias}=0.5\hbar\Gamma$ curve $(2)$, $eV_{bias}=1.0\hbar\Gamma$
curve $(3)$, $eV_{bias}=1.5\hbar\Gamma$ curve $(4)$. Solid (red
online) and short-dashed (blue online) portions of each curve
indicate resonance frequency values with positive and negative
current change $\Delta I$, respectively. Inset: electronic
occupation at resonance frequency against effective gate voltage
(shifted of $E_{p}$) for $eV_{bias}=0.1\hbar\Gamma$ (curve $(1)$)
and $eV_{bias}=1.5\hbar\Gamma$ (curve $(4)$). In this plot
$A_{ext}=10^{-3}$, $\omega_{0}/\Gamma=0.01$ and
$E_{p}/\hbar\Gamma=k_{B}T/\hbar\Gamma=0.1$.}\label{fig1}
\end{figure}

The temperature dependence of the current-frequency profiles,
observed in the experiments, exhibits a nontrivial behavior that
supports our model. As shown in Fig.\ref{fig3}a, for very small
temperatures, a triangular shape is found as discussed above. On
the other hand, for sufficiently large temperatures, the
current-frequency profile turns into Lorentzian shape,
characteristic of the linear response regime (Fig.\ref{fig3}c).
This counterintuitive behavior is determined by a significative
reduction of the intrinsic nonlinear terms in the effective force
$F(x)$ on the resonator as function of the temperature. Actually,
the correction term due the average electronic occupation (eq.
(\ref{occu})) tends to become independent of the resonator
displacement $x$. Furthermore, the broadening of the
current-frequency profiles as function of the temperature is
produced not only by this effect but also by the growth of the
intrinsic damping coefficient $A(x)$ in the Langevin equation
eq.(\ref{Langevin1}). Indeed, with increasing temperatures, $A(x)$
increases where resonator dynamics occurs. The temperature
dependence of the intrinsic damping $A(x)$ is also responsible for
the behavior of the resonator quality factor $Q$, defined as
$I_{0}/\Delta I_{half-high}$. In the regime where the
current-frequency profiles exhibit a Lorentzian shape, we find a
power-law of the kind $T^{-0.40}\;$ \cite{Nota3} that is quite
close to that found in the experiment ($T^{-0.36}$). However, it
is important to point out that this exponent depends on the gate
voltage effectively applied to the device.

Finally, we have analyzed all the traces of current variations as
function of the antenna frequency, obtained tuning the effective
gate voltage. Going from a low to a high current-carrying state,
the characteristic dip of the resonance frequency as function of
the effective gate voltage is obtained in excellent agreement with
experiments (curve (1) of Fig.\ref{fig1}). The observed
renormalization of the resonance frequency can be related to
strong variations of the electronic occupation eq.(\ref{occu}) as
function of the gate voltage (see solid line in the inset of
Fig.\ref{fig1}). When the device is in a low current-carrying
state ($|V_{gate}^{eff}-E_{p}|>1.5\hbar\Gamma$), the average
electronic occupation is not sensitive to gate voltage variations.
Instead, in a high current-carrying state
($|V_{gate}^{eff}-E_{p}|<1.5\hbar\Gamma$), the electronic
occupation shows a strong variation, providing the softening of
the resonance frequencies. Increasing the bias voltage to values
closer to $\hbar\Gamma$ or larger (line (3) and (4) of Fig.
\ref{fig1}), one obtains a broadening of the resonance frequency
dip, owing to a wider conduction window with respect to the
broadening ($\sim \hbar\Gamma$) of the electronic energy level.
Actually, with increasing the bias, the electronic contribution to
the effective spring constant increases, producing a nontrivial
renormalization of the resonance frequency as function of the
gate. We note that for $eV_{bias}^{eff}=1.5\hbar\Gamma$ (line (4)
of Fig. \ref{fig1}), a fine structure represented by two very
small dips appears. When the bias window, whose extension is
proportional to $eV_{bias}^{eff}$, becomes larger than the
broadening of the level, one could tune the electronic device into
a region of conducting states where the variations of the
occupation are smaller than that obtained at the boundary of the
conduction window itself (see dashed line in the inset of
Fig.\ref{fig1}). When the electronic device goes through states
with different conducting character, the maximum renormalization
of the resonance frequency occurs, providing two dips in the
resonance frequency of the nanotube. This feature could be
experimentally observed with a larger resolution in the applied
gate voltage, when a very large bias is applied to the nanotube.

\section{Conclusions}
In conclusion, we have studied a self-detecting SET realized by a
suspended carbon nanotube including, in a non-perturbative way,
the effect of the antenna driving the nanotube toward a nonlinear
regime. All the qualitative features of the device, experimentally
observed, are accurately reproduced clarifying the origin of the
nonlinear effects. The nonlinear behavior is understood without
adding \emph{by hand} nonlinear terms to the effective force
exerted on the resonator \cite{Steele,Weick,Weick1,Rama}, but
stems out naturally from the nontrivial nonequilibrium
time-dependent electronic occupation controlled by the coupling
with the leads. We have shown that, increasing the temperature,
the nonlinear effects in the current-frequency response are washed
out as a result of the increase of the intrinsic damping of the
resonator and of the reduction of the intrinsic nonlinear terms of
the effective self-consistent force. Within our approach, a
broadening of the frequency dip as function of the bias voltage is
reproduced, predicting in the limit of large bias a double dip
structure.

\section{ACKNOWLEDGMENTS} A.Nocera acknowledges G. A. Steele for
very useful clarifications and CNISM for the financial support.

\appendix
\section{}
Here, we briefly show that the model Hamiltonian for the vibrating
nanotube encapsulated in the eq.(\ref{Htot}) is equivalent to that
used in \cite{Steele,Huttel} and in other papers in the literature
\cite{Labadze,Blanter,Blanter1,Bennett,Weick,genovesi}. Indeed, as
clearly discussed in Ref. \onlinecite{Clerk}, a single electronic
transistor (SET) consists of a metallic dot (represented by a
nanotube in the case considered in Ref.
\onlinecite{Steele,Huttel}) with a large Coulomb-charging energy
$E_{C} = e^{2}/2C_{tot}$ ($C_{tot}$ is the total capacitance of
the dot) coupled via tunnel junctions to both a source and a drain
metal electrode. The Hamiltonian for the electronic and
vibrational degrees of freedom of the dot is given by
\begin{equation}\label{ChargeH}
H_{dot} = E_{C}(N_{el} - N_{gate})^{2}+ {1\over 2}kx^{2},
\end{equation}
comprising a charging-energy term and the harmonic vibrational
energy. In eq.(\ref{ChargeH}), $N_{el}$ is the charge on the SET
dot and $N_{gate} = C_{gate}V_{gate}/e$ is the dimensionless
electron number associated with a gate voltage $V_{gate}$ which is
coupled to the dot via a capacitance $C_{gate}$. In addition, a
voltage $V_{bias}$ is applied between source and drain which
drives the tunneling of electrons across the SET. Here,
$C_{tot}=C_{gate}+C_{leads}$, where $C_{leads}$ is the sum of the
capacitances resulting from the coupling to the leads. When the
nanotube is allowed to vibrate, the gate capacitance $C_{gate}$
assumes a spatial dependence $C_{gate}(h(x))$, where $h$ is the
distance between the nanotube and the gate electrode when the
nanotube is displaced by a distance $x$ from its equilibrium
position ($h(x)=h_{0}+x$). In the limit of small displacements,
one can expand $\hat H_{dot}(x)$ around $x=0$ obtaining a
Holstein-like linear correction term in $x$:
\begin{equation}\label{Holcoupling}
H_{int}=\lambda N_{el} x,
\end{equation}

where $\lambda=-2(E_{C}V_{g}/e)(dC_{g}/dx)$. In the small energy
window of interest for a single dip feature investigated in Ref.
\onlinecite{Steele,Huttel}, we can neglect the weak gate and bias
voltage dependence of $\lambda$ that is assumed constant
\cite{Bennett,Clerk,Labarthe}. Moreover, if we choose $x=0$ as the
position where the Coulomb force for $N_{el}=0$ electrons in the
island equals the elastic force, $\lambda$ can be then interpreted
as the net force acting on the nanotube when one excess electron
is populating the nanotube itself. The terms independent on
$N_{el}$ in the above expansion are usually neglected
\cite{Clerk,Labarthe}. One further assumes that the gate voltage
is such that only charge states with $N_{el}=0$ and $1$ are
accessible. In this case, $N_{el}^{2}=N_{el}$ and one can
incorporate remaining constant terms (independent on $x$) in the
above expansion in an effective gate voltage obtaining
\begin{equation}\label{Hdev}
H_{dot}\simeq V_{gate}^{eff}N_{el}+\lambda N_{el} x + {1\over
2}kx^{2}.
\end{equation}
If we quantize the electronic and vibrational degrees of freedom,
eq.(\ref{Holcoupling}) gives the Holstein coupling discussed in
the main text.


\begin{thebibliography}{99}
\bibitem{Nota0} Usually, increasing the temperature, one would
expect to access the region far away the potential minimum, where
nonlinear effects could be present.
\bibitem{Nota000} For the sake of simplicity, back-actions effects
of the vibrating resonator on the electromagnetic field of the
antenna are disregarded.
\bibitem{Nota00} In sequential tunneling regime, $\hbar\Gamma << min[eV_{bias},k_{B}T]$,
the tunneling events between the leads and the molecule are
energy-conserving, with the rates determined by Fermi's golden
rule at the lowest order in perturbation theory in the tunneling
Hamiltonian $\hat H_{tun}=\sum_{k,\alpha}(V_{k,\alpha}{\hat
c^{\dag}_{k,\alpha}}{\hat d}+ h.c.)$. In the cotunneling regime,
$eV_{bias} < \hbar\Gamma < k_{B}T$ or $k_{B}T < \hbar\Gamma <
eV_{bias}$ or $max[eV_{bias},k_{B}T]< \hbar\Gamma$, higher-order
tunneling events become dominant.
\bibitem{Nota2} It is very difficult to infer the actual values of the effective
source-drain bias $eV_{bias}^{eff}$ and the gate voltage
$V_{gate}^{eff}$ applied to the device from the experimental
setup. Due to the capacitive coupling between the carbon nanotube
and the surrounding electrodes and to the length ($400-1100nm$) of
the nanotube itself, only a fraction of the applied bias and gate
voltage acts effectively onto the nanotube level interested in the
transport \cite{Steele,Huttel}.
\bibitem{Nota1} We consider for simplicity spin-less electrons, as the inclusion
of spin as well as on-site Coulomb repulsion doesn't qualitatively
change our results.
\bibitem{Nota3} We have verified that, in the range of temperature
investigated ($0.05<k_{B}T/\hbar\Gamma<1$ $\rightarrow$
$60mK<T<1.2K$), the the widths of current frequency profiles are
indipendent on the external antenna amplitude applied to tune the
resonator in the linear response regime.
\end{thebibliography}
\end{document}